\documentclass[iicol,sn-basic]{sn-jnl}
\usepackage{graphicx}
\graphicspath{ {./images/} }
\usepackage{esvect}
\usepackage{amsmath}
\usepackage[utf8]{inputenc}
\usepackage[tight,footnotesize]{subfigure}
\usepackage{appendix}
\usepackage{lscape}
\usepackage{textcomp}
\usepackage{amssymb}
\usepackage{lscape}
\usepackage{bbm}
\usepackage{hhline}
\usepackage{multirow}
\usepackage{adjustbox}
\usepackage{makecell}
\usepackage{subcaption} 

\jyear{2024}%

\theoremstyle{thmstyleone}%
\theoremstyle{thmstyletwo}%

\theoremstyle{thmstylethree}%

\begin{document}

\title{SincPD: An Explainable Method based on Sinc Filters to Diagnose Parkinson's Disease Severity by Gait Cycle Analysis}


\author*[1]{\fnm{Armin} \sur{Salimi-Badr}}\email{a\_salimibadr$@$sbu.ac.ir}

\author[1]{\fnm{Mahan} \sur{Veisi}}

\author[1]{\fnm{Sadra} \sur{Berangi}}


\affil*[1]{\orgdiv{Faculty of Computer Science and Engineering}, \orgname{Shahid Beheshti University},  \city{Tehran}, \country{Iran}}



\abstract{In this paper, an explainable deep learning-based classifier based on adaptive sinc filters for Parkinson's Disease diagnosis (PD) along with determining its severity, based on analyzing the gait cycle (SincPD) is presented. Considering the effects of PD on the gait cycle of patients, the proposed method utilizes raw data in the form of vertical Ground Reaction Force (vGRF) measured by wearable sensors placed in soles of subjects' shoes. The proposed method consists of Sinc layers that model adaptive bandpass filters to extract important frequency-bands in gait cycle of patients along with healthy subjects. Therefore, by considering these frequencies, the reasons behind the classification a person as a patient or healthy can be explained. In this method, after applying some preprocessing processes, a large model equipped with many filters is first trained. Next, to prune the extra units and reach a more explainable and parsimonious structure, the extracted filters are clusters based on their cut-off frequencies using a centroid-based clustering approach. Afterward, the medoids of the extracted clusters are considered as the final filters. Therefore, only 15 bandpass filters for each sensor are derived to classify patients and healthy subjects. Finally, the most effective filters along with the sensors are determined by comparing the energy of each filter encountering patients and healthy subjects. 
}




\maketitle

\section{Introduction}

Parkinson's Disease (PD) is the second most common neurodegenerative disease affecting many elderly individuals worldwide \citep{neuro-fuzzy,Liu2021CNN_and_LSTM,khoury2019data,pan2012parkinson,GOYAL2020103955}. It primarily originates from the loss of dopaminergic neurons in the Substantia Nigra pars Compacta within the Basal Ganglia \citep{Hall2020,salimi2017possible,salimi2018system}.

PD treatment typically begins after symptoms manifest years post-infection, necessitating more complex treatments like Deep Brain Stimulus (DBP) instead of simpler lifestyle changes. Early diagnosis can enable more effective and economical treatments. However, diagnosis is a challenging task that prompts the use of machine learning to enhance healthcare diagnostics.

PD patients often exhibit movement issues such as Bradykinesia, Akinesia, stiffness, and resting tremor \citep{Hall2020}. Consequently, machine learning methods focus on movement analysis for diagnosis, including \textit{speech disorders} \citep{Kuresan2021,Pramanik2022,Yousif2023-gy,Liu2022}, \textit{g`ait changes} \citep{khoury2018cdtw,salimipd,deep1d,LSTM,SalimiBadr-MultiLSTM,Liu2021CNN_and_LSTM}, \textit{handwritten records} \citep{Yousif2023-gy}, \textit{typing speed} \citep{prashanth2016high}, \textit{eye movement changes} \citep{Farashi2021}, multi-modal biomedical time-series analysis \citep{JUNAID2023107495}, and postural stability assessment using RGB-Depth cameras \citep{Ferraris2024-up}.

Machine Learning, particularly Deep Learning, has been effectively applied in medical diagnosis \citep{deep1d,cnnlstm,Liu2021CNN_and_LSTM}. Explainable methods like neuro-fuzzy systems \citep{neuro-fuzzy} are limited by their reliance on high-level clinical features. Deep models extract complex features but lack interpretability \citep{xai1,sensors1}. However, explainability is crucial in medical applications to support expert diagnosis \citep{JUNAID2023107495,Saadatinia2024}.

Convolutional Neural Networks (CNNs) extract abstract features through learned filters. The first convolutional layers are critical as they process raw signals and form higher-level features. \textit{SincNet} \citep{SincNet} enhances interpretability by using sinc-shaped bandpass filters with only two parameters: center and width. This allows identification of important frequency bands influencing network decisions \citep{hung2022using,ravanelli2018interpretable}.

This paper presents an explainable AI approach to detect PD and assess its severity using sinc-layers in CNNs to analyze vertical-Ground Reaction Force (vGRF) signals from wearable sensors (SincPD). The Sinc layers model adaptive bandpass filters to extract key frequency bands in gait cycles, enabling interpretation of the network's decisions.

Our SincPD involves preprocessing, training a large filter-rich model, pruning redundant filters via centroid-based clustering, and selecting cluster medoids as final filters. We analyze important frequencies by calculating filter energy for patients and healthy subjects.

The paper is organized as follows: Section \ref{sec3} reviews related machine learning studies for PD detection. Section \ref{sec2} presents preliminaries. Section \ref{sec4} details the proposed methodology, including preprocessing, SincNet architecture, learning, and pruning. Section \ref{sec5} discusses experimental results and filter analysis. Conclusions are in Section \ref{sec6}.

\section{Related Work}
\label{sec3}

Parkinson's Disease (PD) is a progressive neurodegenerative disorder marked by motor symptoms (e.g., tremors, rigidity) and non-motor manifestations. Due to the movement disorders, gait analysis have been one of the most popular approaches to study the Parkinson's disease in the literature. 

Gait analysis has proven valuable for PD diagnosis due to its ability to identify changes such as reduced stride length, slower speed, and increased variability. \cite{pistacchi2017gait} observed significant differences in cadence, stride and stance duration, and gait velocity in early-stage PD patients compared to controls. \cite{SOFUWA20051007} reported shorter step length and slower walking speed, while \cite{lescano2016possible} found deviations in stance and swing phases and ground reaction force at Hoehn and Yahr stages 2--2.5.

Numerous sensor-based systems have been developed for automated PD detection via gait analysis, extensively utilizing vGRF captured by in-shoe sensors~\citep{deep1d,LSTM,SalimiBadr-MultiLSTM,Applied_Soft_SVM,Applied_Soft_severity,cnnlstm,Liu2021CNN_and_LSTM}.\cite{Applied_Soft_SVM} employed spatiotemporal features with a multi-class support vector machine. \cite{neuro-fuzzy} presented a type-2 fuzzy logic approach using vGRF data. and \cite{deep1d} proposed a 1D convolutional neural network for PD detection and severity estimation. \cite{ZHAO201891} utilized a dual-channel LSTM for gait classification, though limited by partial gait acquisition. \cite{LSTM} applied LSTM networks for PD diagnosis and severity rating without hand-crafted features, and \cite{BVIDYA2022105099} introduced a CNN-LSTM model based on empirical mode decomposition of vGRF signals.

Although these machine learning methods have advanced PD diagnosis, many depend on hand-crafted features or lack interpretability.
 To overcome these limitations, we propose an explainable deep learning model using Sinc filters in a convolutional neural network (CNN) to extract salient frequency bands from raw vGRF data. This approach combines high classification accuracy with transparency, improving the clinical relevance of PD diagnosis and severity assessment.

\section{Preliminaries}
\label{sec2}

In this section, we review the preliminaries of the proposed method, including the concept of bandpass filters along with the neural architecture based on them (SincNet).




\noindent\textbf{SincNet Model Architecture.} \label{subsec:sincnet}SincNet introduces a novel approach by incorporating parameterized sinc functions into a convolutional neural network (CNN) architecture to efficiently create bandpass filters. This method allows SincNet to learn from only two parameters per filter—lower and upper cutoff frequencies \( f_1 \) and \( f_2 \)—significantly reducing model complexity.

\begin{figure}[ht]
    \centering
    \begin{minipage}{0.493\columnwidth}
        \centering
        \includegraphics[width=\textwidth]{ 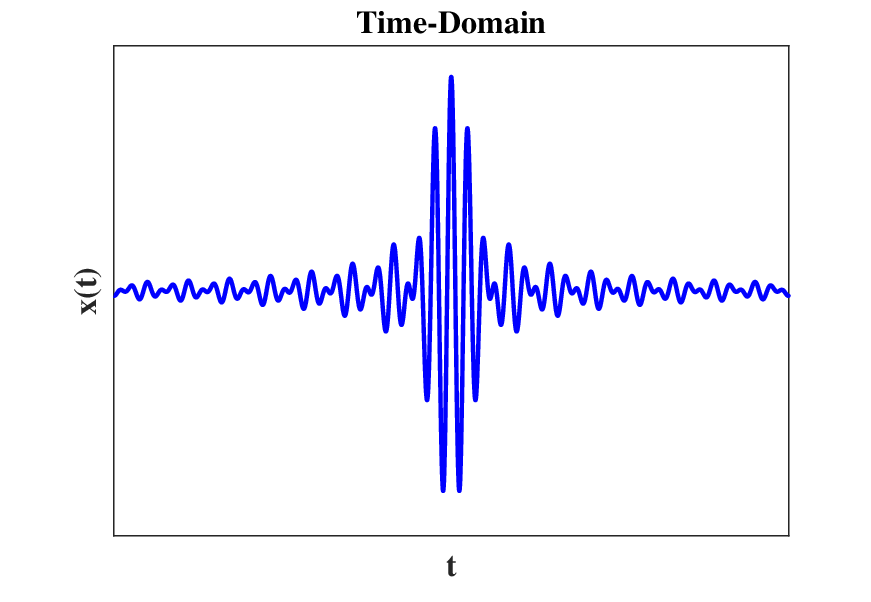}
    \end{minipage}
    \hfill
    \begin{minipage}{0.493\columnwidth}
        \centering
        \includegraphics[width=\textwidth]{ 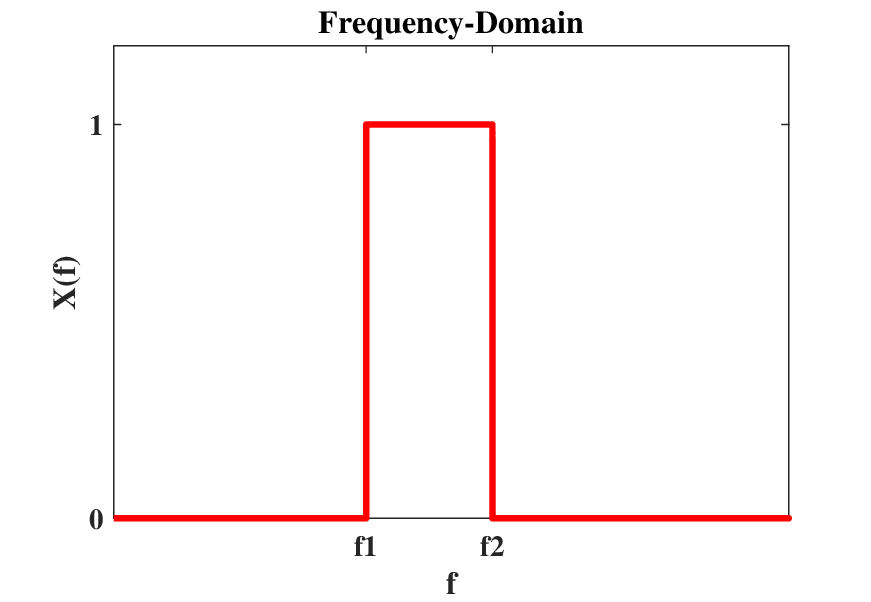}
    \end{minipage}
    \caption{The impulse-response along with the frequency-response of a bandpass filter that allows selective frequency passage from \( f_1 \) to \( f_2 \).}
    \label{fig:sinc}
\end{figure}

\noindent\textbf{Bandpass Filter Configuration.} In SincNet, each filter’s impulse response is modeled by the difference between two scaled sinc functions, representing a bandpass filter \citep{SincNet}:

\begin{equation}
h[n] = 2f_2 \cdot \text{sinc}(2\pi f_2 n) - 2f_1 \cdot \text{sinc}(2\pi f_1 n),
\end{equation}

where the sinc function is defined as follows:

\begin{equation}
\text{sinc}(\omega) = 
\begin{cases} 
\frac{\sin(\omega)}{\omega} & \text{if } \omega \neq 0, \\
1 & \text{if } \omega = 0.
\end{cases}
\end{equation}

This impulse response \( h[n] \) allows selective frequency passage from \( f_1 \) to \( f_2 \), effectively attenuating frequencies outside this range. The form of a sample of this impulse response along with its frequency response is presented in Fig.~\ref{fig:sinc}. The design is inspired by the properties of sinc, which acts as an ideal mathematical model for bandpass filters due to its sharp frequency domain characteristics.

\noindent\textbf{Frequency-Domain Characterization.} The frequency response of the bandpass filter designed by SincNet is fundamentally a rectangular function (see Fig.~\ref{fig:sinc}), characterized as:
\begin{equation}
G(f) = \text{rect}\left(\frac{f}{2f_2}\right) - \text{rect}\left(\frac{f}{2f_1}\right),
\end{equation}
where \( \text{rect}(\cdot) \) is the rectangular function. The equivalent time-domain representation is shown in Equation~(4), illustrating how the sinc functions are used to implement bandpass behavior. The filters are initialized to cover a range of frequencies between 0 and half the sampling rate (\( f_s/2 \)), guided by psychoacoustic principles such as the equivalent rectangular bandwidth (ERB), which aids in setting the filters' bandwidth more precisely.

\section{Materials and Methods}
\label{sec4}

In this section, we detail the methodology employed in our study, encompassing the preprocessing steps, the architecture of the proposed deep learning model, and the pruning process.

\subsection{Preprocessing}
\label{sec42}

The preprocessing stage ensures input data quality and suitability for the neural network. We prepared the dataset by partitioning it into ten-second segments, filtering out incomplete or inconsistent data, and standardizing the signals. This process ensured clean and normalized inputs for further analysis.


To reduce dimensionality without data loss, we calculated the difference between left and right sensor signals. This approach reduced 16 arrays per subject to 8, effectively preserving critical gait patterns while capturing meaningful inter-sensor variations. 

\begin{figure}
    \centering
    \includegraphics[width=3.in]{ 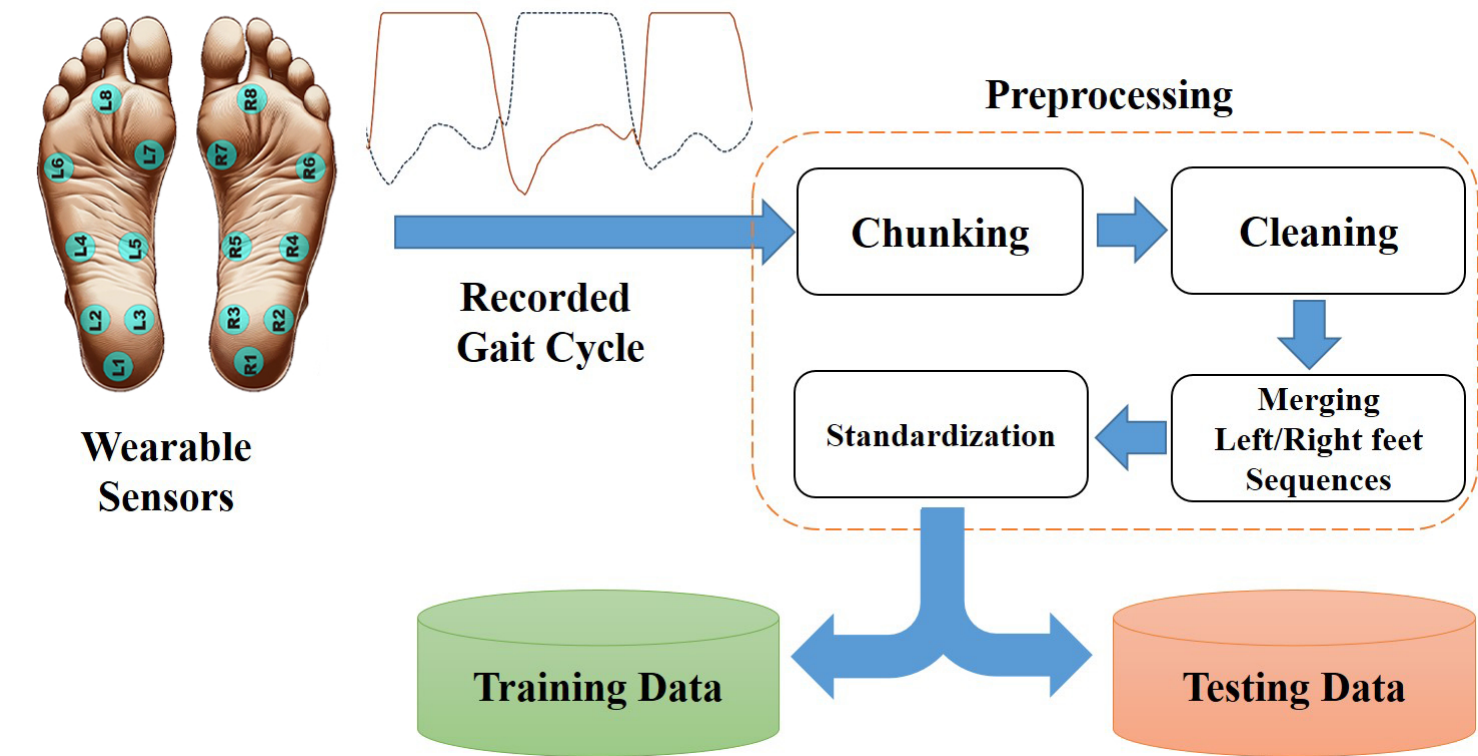}
    \caption{Overview of the preprocessing pipeline.}
    \label{fig:pre}
\end{figure}

Moreover, stratified splitting was employed to maintain class balance in the training and test datasets, ensuring robustness across patient and healthy subject categories. In summary, Fig.~\ref{fig:pre} demonstrates an overview of the preprocessing pipeline, including chunking, cleaning, and standardization steps.


\subsection{Proposed Architecture}
\label{sec43}

\begin{figure*}[t]
    \centering
    \includegraphics[width=0.9\textwidth]{ 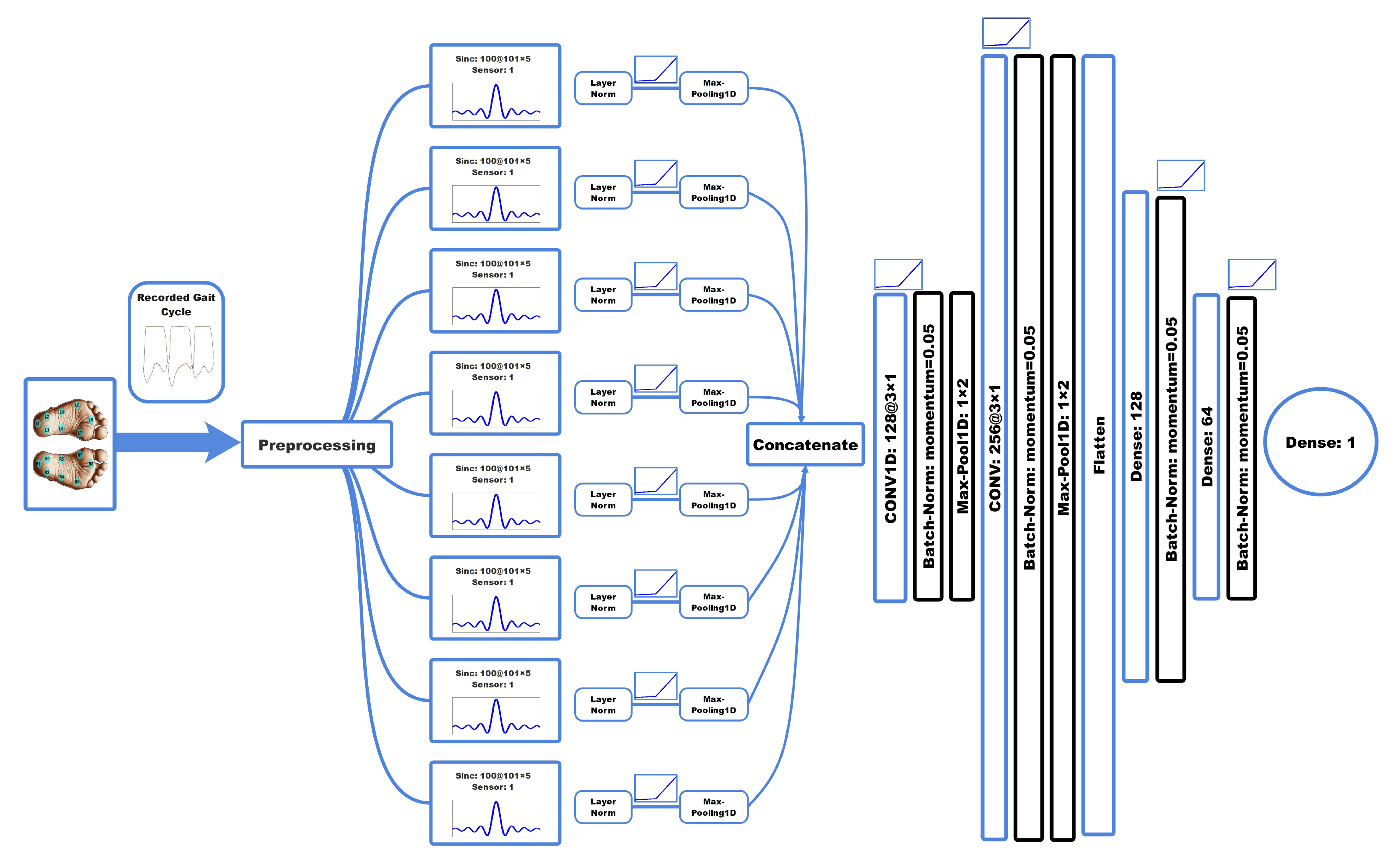}
    \caption{Architecture of the SincPD. The model begins with an input layer followed by SincConv1D layers, specialized for handling 1D signals. These layers are normalized and activated using the Leaky ReLU function. Following the SincConv1D layers, traditional convolutional layers with batch normalization extract higher-level features. Finally, dense layers are utilized for classification. }
    \label{fig:model_architecture}
\end{figure*}

The initial model architecture, depicted in Fig.~\ref{fig:model_architecture}, is designed to efficiently extract and refine meaningful features from vGRF data using a sequential framework. Traditional convolutional layers, while powerful for feature extraction, often require a high number of learnable parameters and lack mathematical interpretability, making them less suitable for tasks prioritizing explainability and efficiency.

To address these challenges, the first layers of our model utilize SincConv1D filters. These filters are particularly effective in capturing critical low-level features essential for subsequent processing while maintaining interpretability due to their frequency domain meaning (adaptive bandpass filters). Each filter is parameterized by only two learnable parameters (cutoff frequencies), reducing model complexity and allowing direct analysis of the learned frequency bands. The subsequent layers and their configurations, described in this section, refine these features into robust predictive insights.

\subsubsection{SincConv1D Layers}
\label{subsubsec:sincconv1d_layers}

We employ eight SincConv1D layers (one per preprocessed signal) to extract frequency-specific features from vGRF data. In the initial version of the model, each layer is configured with 100 filters of length 101. The extracted features are normalized, activated with Leaky ReLU, and pooled to reduce computational load. Finally, all eight outputs are concatenated, forming a unified representation for subsequent network layers.

\subsubsection{Model Structure }
\label{subsubsec:conv_dense_combined}

Following the frequency-focused SincConv1D feature extraction, two standard Conv1D layers (128 and 256 filters) further refine the learned representations. These layers are followed by Batch Normalization, Leaky ReLU, Dropout, and Max Pooling to improve model generalization and computational efficiency.

The outputs are then flattened and passed through three Dense layers, with 128, 64, and 1 neuron(s), respectively. The first two layers use ReLU activation, while the final layer employs a sigmoid activation function to output a probability score between 0 and 1. Batch Normalization and L2 regularization are applied to enhance stability and reduce overfitting. Finally, the Adam optimizer~\citep{kingma2014adam} is applied to learn model's parameters by minimizing the cross-entropy.

    \subsection{Pruning Method}
    
        High number of filters decreases the interpretability. Therefore, we propose a pruning method to prune extra filters. To realize this, we cluster filters based on their cut-off frequencies and reconstruct the architecture based on the clusters' centroids.
        
        As previously described~\ref{subsec:sincnet}, SincConv1D layers learn the center frequency ($f_c$) and bandwidth ($b$) of sinc filters, which define their frequency response. These parameters are used for clustering and identifying redundancy. Using K-means clustering~\citep{kmeans}, we group similar ($f_c$, $b$) parameters and retain only the most significant clusters.
        
        To determine the optimal number of clusters, we apply the elbow method and silhouette scores~\citep{geron2022hands}, initially optimizing for Sensor~1. For each sensor, the optimal $k$ value is identified by analyzing silhouette diagrams for Sinc layers. The clusters and centroids are then visualized to ensure the retained filters capture critical patterns.
        
        The new architecture is constructed using the cluster centroids as filter weights in the SincConv1D layers, reducing parameters while preserving essential patterns. The model is retrained for a few epochs to fine-tune performance, with the rest of the architecture unchanged.

    \section{Experimental Results}
    \label{sec5}
    In this section, the performance of the proposed method is evaluated and compared to some previous methods in both diagnosis and severity determination. All experiments have been conducted in CoLab runtime environment based on Python, using the TensorFlow platform along with the SKLearn package.
    
    \subsection{Data and Evaluation Metrics}
    \label{subsec:Data_Evaluation}
    
\begin{figure}
        \centering
        \includegraphics[height= 1.5in,width=3in]{ 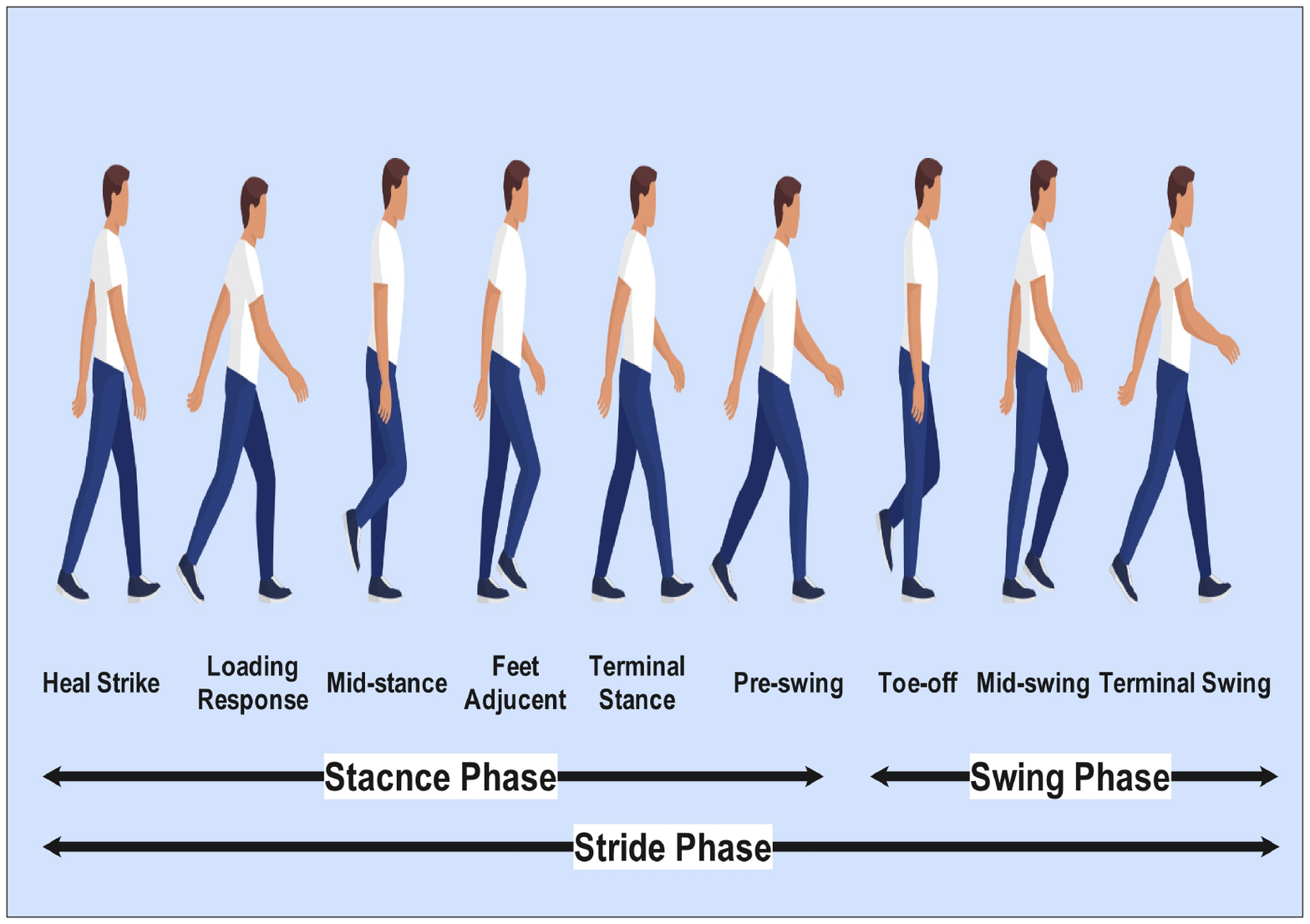}
        \caption{Illustration of the gait cycle phases. The gait cycle consists of the stance phase (heel strike, loading response, mid-stance, terminal stance, pre-swing includes $60\%$ of the cycle) and the swing phase (toe-off, mid-swing, terminal swing include $40\%$ of the cycle). }
        \label{gate_phases_fig}
    \end{figure}

    This study employs gait cycle data from PhysioNet\footnote{https://physionet.org/content/gaitpdb/1.0.0/}, comprising vGRF signals recorded via 16 sensors under participants’ feet. The dataset includes individuals with PD and healthy controls, capturing approximately two minutes of walking at a self-selected pace. Gait cycles (Fig.~\ref{gate_phases_fig}), divided into stance and swing  phases, highlight PD-related alterations in stride length and variability.
    
    A total of 166 subjects (93 PD, 73 controls) participated, with data aggregated from three separate studies (Table~\ref{tab:dataset}). PD severity was assessed based on the modified Hoehn and Yahr scale (Table~\ref{tab:PDseverity}), ranging from Stage~1 (unilateral involvement) to Stage~5 (complete disability).
    
    To evaluate the proposed approach, standard classification metrics including Accuracy, Precision, Recall, and F1 Score were employed.

    \begin{table}
        \centering
        \caption{Number of participants in datasets categorized by severity level.}
        \captionsetup{font={footnotesize,rm}}

        \label{tab:dataset}
        \footnotesize
        \begin{tabular}{|c|c|c|c|c|}
        \hline
        \textbf{Dataset} & \textbf{Healthy} & \textbf{Stage 2} & \textbf{Stage 2.5} & \textbf{Stage 3} \\
        \hline
        Ga & 18 & 15 & 8 & 6 \\
        Ju & 26 & 12 & 13 & 4 \\
        Si & 29 & 29 & 6 & 0 \\
        \hline
        \end{tabular}
    \end{table}
    
    \begin{table}[tb]
    \renewcommand{\arraystretch}{1.2} 
    \centering
    \captionsetup{font={scriptsize,rm}}

    \caption{PD severity classification based on the modified H\&Y scale.}
    \label{tab:PDseverity}
    \resizebox{\columnwidth}{!}{%
        \begin{tabular}{|c|l|l|}
        \hline
        \textbf{Scale} & \textbf{Description} & \textbf{Stage} \\
        \hline
        1 & One side only & No functional disability \\
        1.5 & One side + axial symptoms & Early stage \\
        2 & Bilateral involvement & No balance impairment \\
        2.5 & Mild bilateral, recovery on pull test & Balance impairment \\
        3 & Mild/moderate bilateral & Impaired postural reflexes \\
        4 & Severe disability & Still capable of walking \\
        5 & Bedridden or wheelchair-bound & Completely disabled \\
        \hline
        \end{tabular}
    }
    \end{table}

    \subsection{Training the Initial Model}
    
\begin{figure}
            \centering
            \includegraphics[width=3in]{ 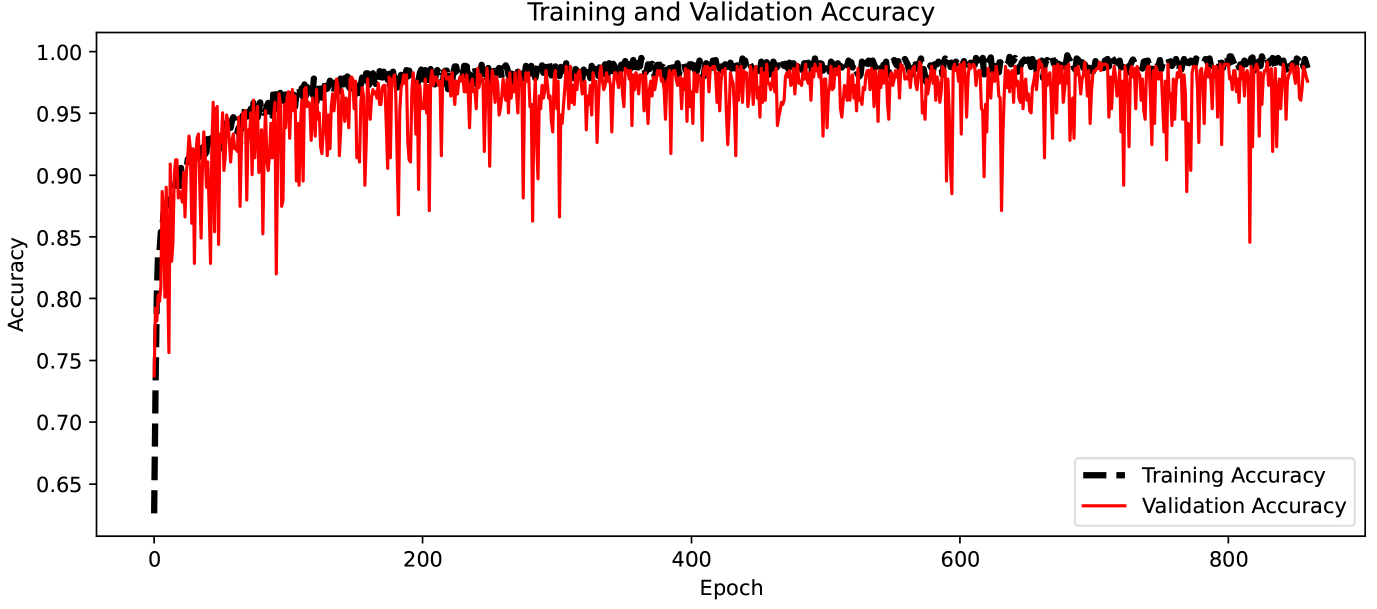}
            \caption {Training and validation accuracy progression over 1000 epochs}
            \label{fig:training_plots}
        \end{figure}

        The explained model in section \ref{sec4} is built and trained using the preprocessed data for 1000 epochs. Training and validation accuracy progression (Fig.~\ref{fig:training_plots}) shows steady improvement, demonstrating effective learning and reliable generalization.
    
        \begin{figure}
            \centering
            \begin{minipage}{0.493\columnwidth}
                \centering
                \includegraphics [width=\textwidth]{ 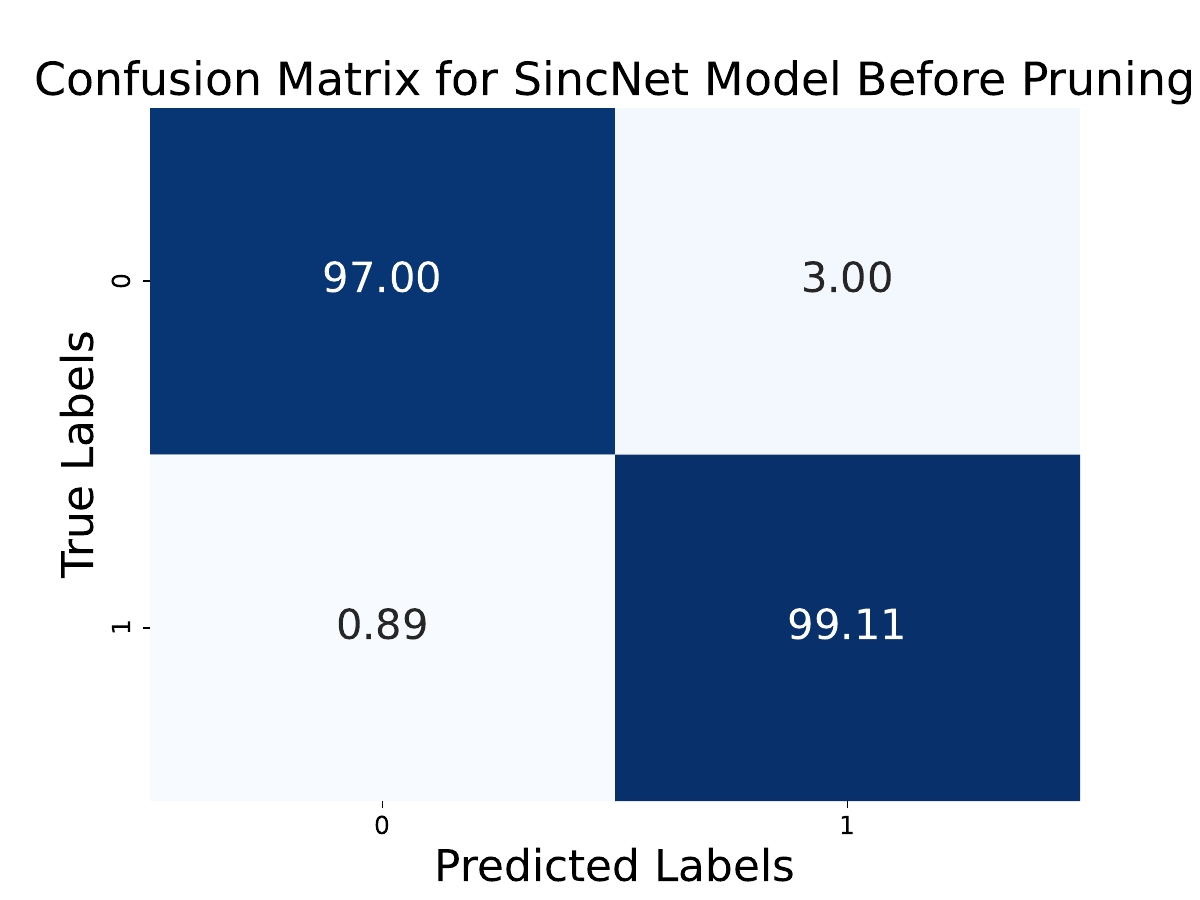}
            \end{minipage}
            \hfill
            \begin{minipage}{0.493\columnwidth}
                \centering
                \includegraphics[width=\textwidth]{ 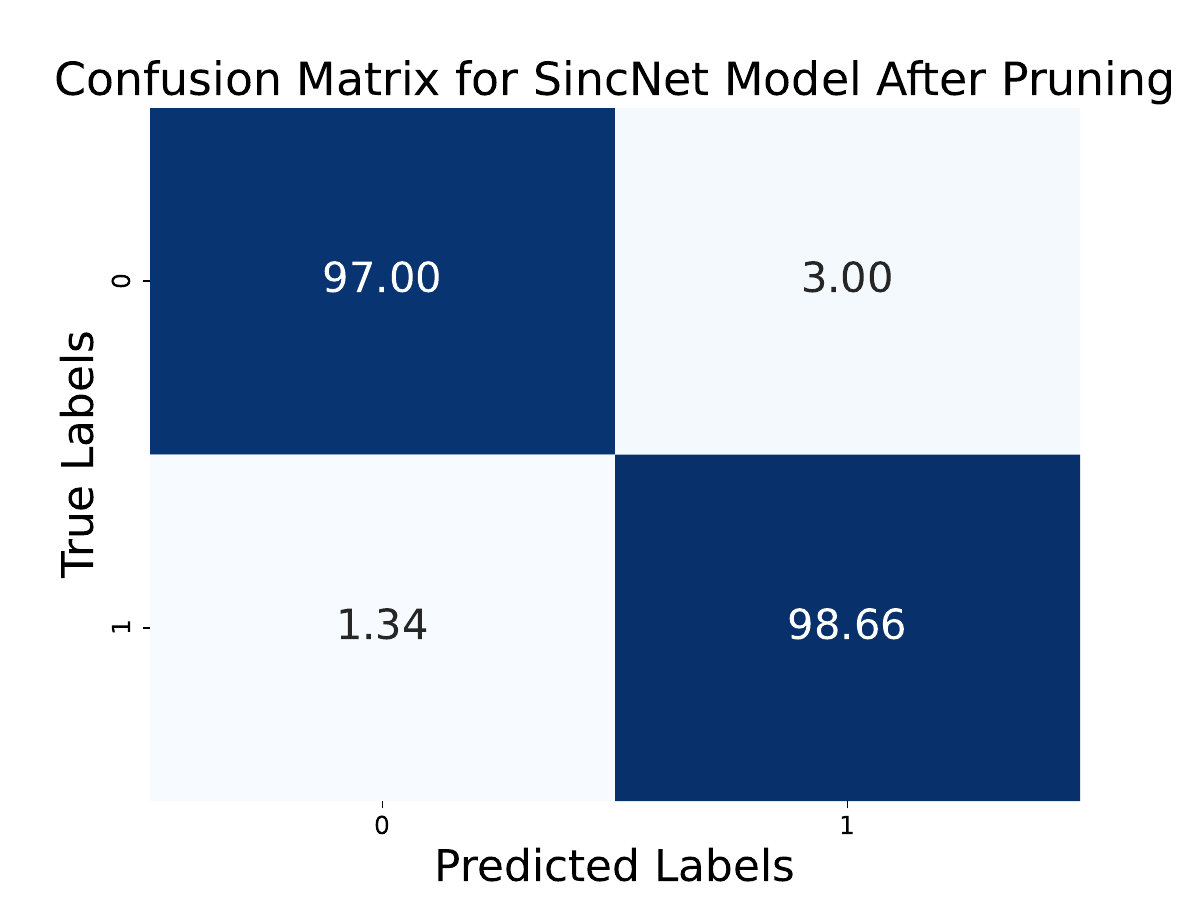}
            \end{minipage}
            \caption{Confusion matrices: Left, classification performance before pruning; Right, classification performance after pruning.}
            \label{fig:combined_confusion}
        \end{figure}

        The performance of the trained SincNet model for binary classification is illustrated through a confusion matrix (Fig.~\ref{fig:combined_confusion}). The model achieves a high classification accuracy, correctly predicting 97\% of samples belong to the negative class and 99.55\% of the positive ones, with minimal misclassifications.
    
    \subsection{Filter Optimization}
    
        To optimize the number of filters in the SincConv1D layers, we utilize the elbow method and silhouette scores, as described in Section~\ref{sec42}, to estimate an initial range for the optimal number of clusters.
    
        \begin{figure}
            \centering
            \begin{minipage}{0.493\columnwidth}
                \centering
                \includegraphics[width=\textwidth]{ 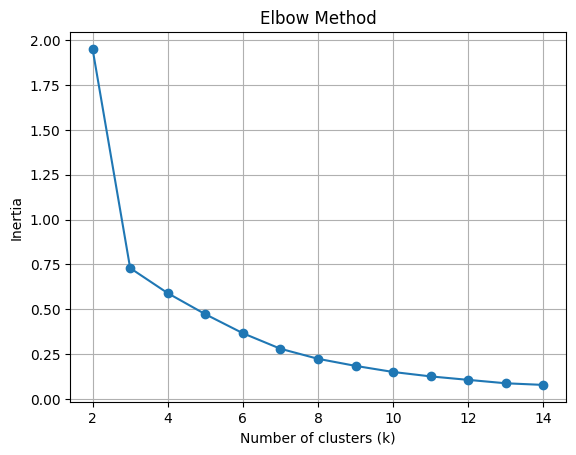}
            \end{minipage}
            \hfill
            \begin{minipage}{0.493\columnwidth}
                \centering
                \includegraphics[width=\textwidth]{ 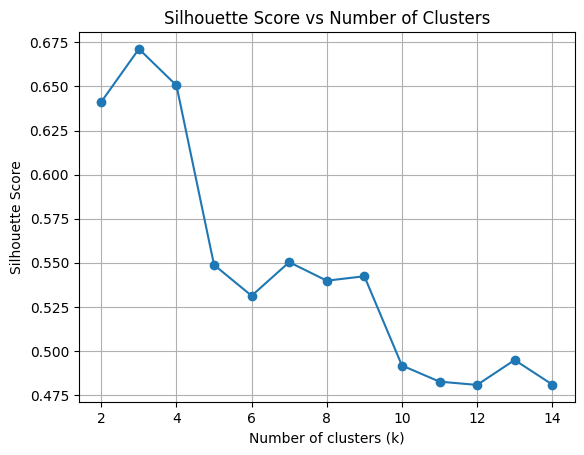}
            \end{minipage}
            \caption{Left: Inertia plot for Sensor 1, using the elbow method to suggest an initial cluster range. Right: Silhouette scores for Sensor 1, peaking around \( k=4 \), confirming well-separated clusters.}
            \label{fig:elbow_silhouette_sensor1}
        \end{figure}
    
\begin{figure}
            \centering
            \includegraphics[width=3in]{ 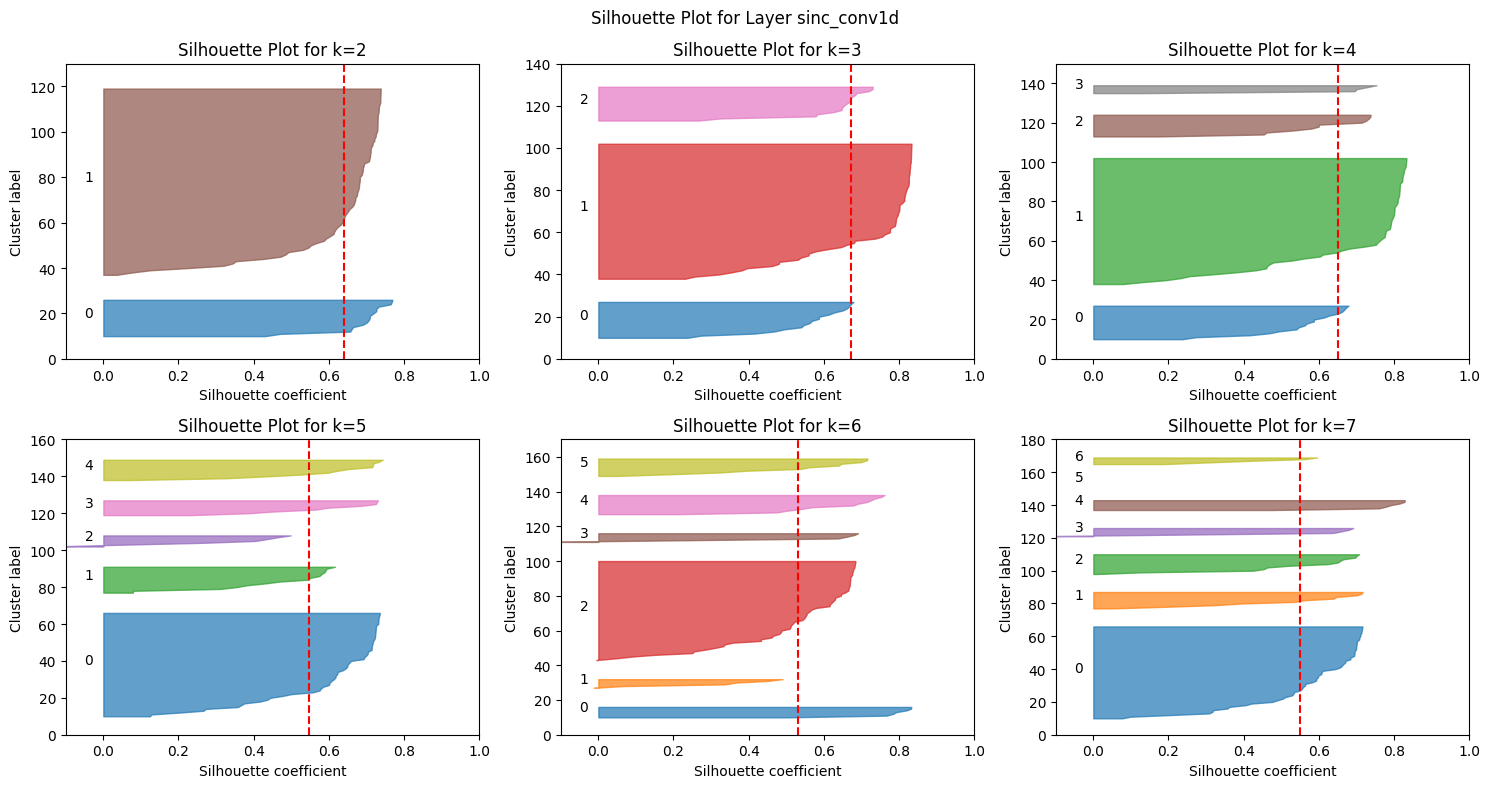}
            \caption{Silhouette Diagram for Sensor 1, visually representing cluster cohesion and separation at the optimal $k$.}
            \label{fig:silhouette_diagram_sensor1}
        \end{figure}

        From the analysis of the plots sampled for Sensor 1 (Fig.~\ref{fig:elbow_silhouette_sensor1}), the elbow method indicates a potential range of 3 to 7 clusters, while the silhouette scores peak around $k=4$. These observations suggest that the optimal number of filters lies within this range, which should be refined and evaluated for each sensor individually. Thus, we analyze the optimal $k$ using the silhouette diagram for each sensor. As shown in Fig.~\ref{fig:silhouette_diagram_sensor1}, the silhouette plots for Sensor 1 suggest $k=3$ as the best choice, providing optimal cluster cohesion and separation. Repeating this procedure for each sensor yields an efficient clustering configuration, improving the overall filter performance.
        
\begin{figure}
            \centering
            \includegraphics[width=1.\linewidth]{ 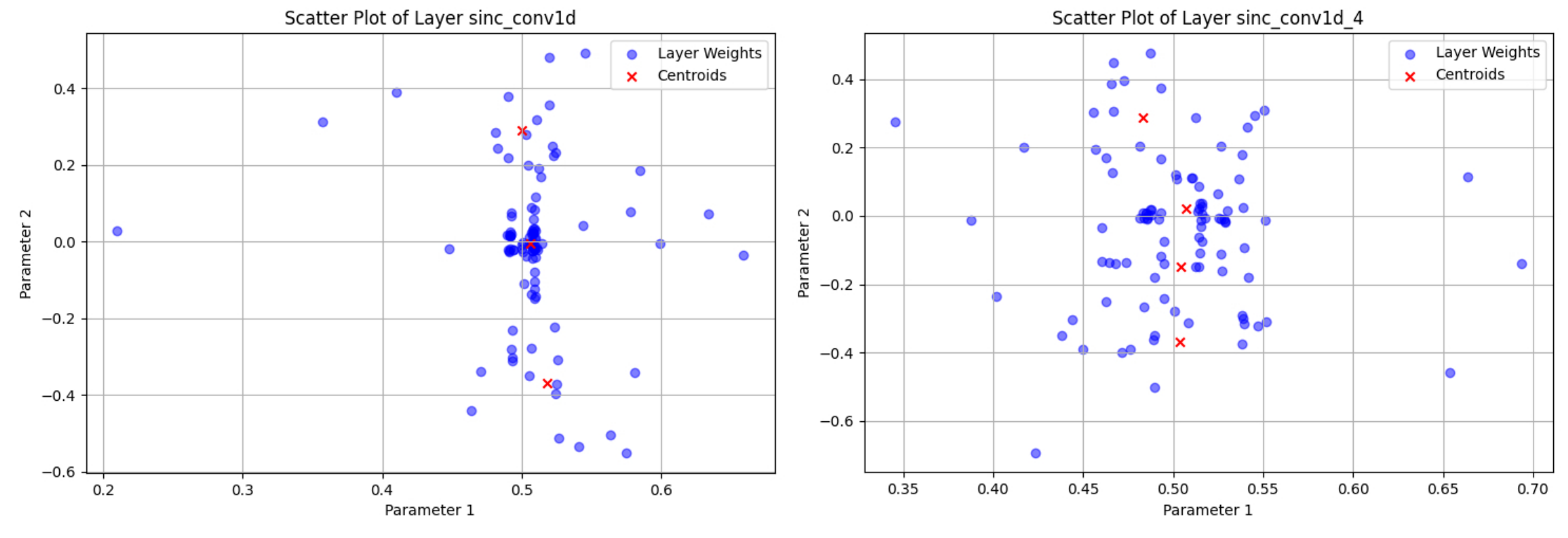}
            \caption{Scatter plots for two sample sensors (Sensor 1 and Sensor 5), showing clustered filters and their centroids post-clustering. The x-axis represents the bandwidth parameter ($b$) and the y-axis represents the center frequency parameter ($f_c$).}
            \label{fig:scatter_plots_all_sensors}
        \end{figure}

        For two sample sensors (Sensor 1 and Sensor 5), scatter plots illustrate the distribution of filter weights and their centroids after clustering. For Sensor 1, the Silhouette Diagram analysis determined $k=3$, and the corresponding centroids were obtained using the k-means algorithm. Similarly, for Sensor 5, $k=4$ was identified as optimal, resulting in four centroid pairs of $(f_c, b)$. Fig.~\ref{fig:scatter_plots_all_sensors} displays these clustered filters and centroids.

        As described in Section~\ref{subsec:sincnet}, after retraining the model with the pruned filters, the updated architecture maintains nearly the same classification performance as the initial model, despite using significantly fewer SincNet filters. The confusion matrix (Fig.~\ref{fig:combined_confusion}, right) shows that the model retains 97\% accuracy for the negative class while experiencing only a minor decrease in accuracy for the positive class, from 99.55\% to 98.66\%.
    
        This minimal reduction in performance is achieved while reducing the total number of filters across the SincNet layers from 800 to approximately 30, significantly decreasing the parameters in the feature extraction layers.

\begin{figure}
            \centering
            \includegraphics[width=3in]{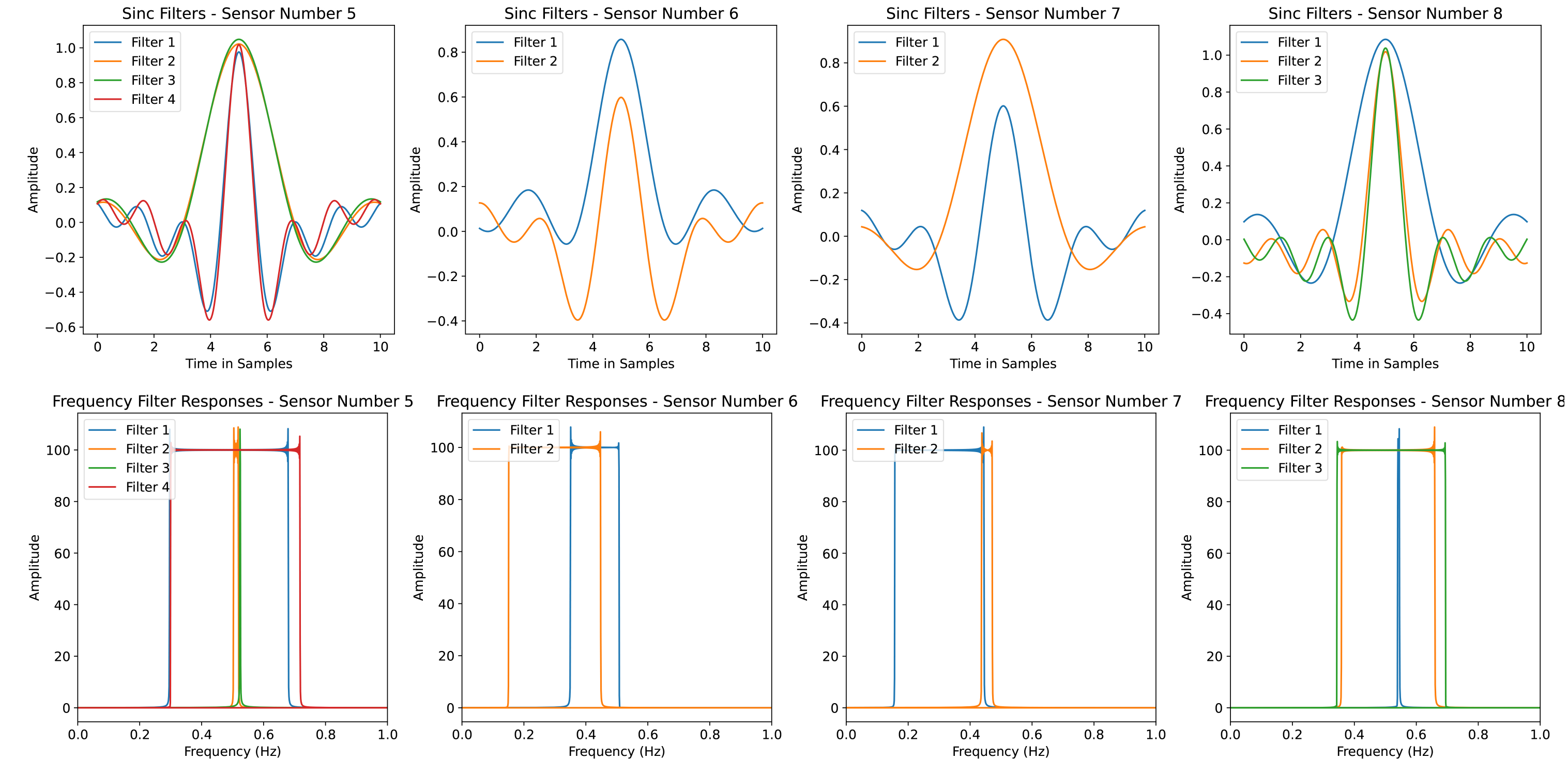}
            \caption{Visualization of Sinc Filters in Different Layers In Time (First Row) and Frequency (Second Row) Domain}
            \label{fig:sinc_filters}
        \end{figure}

        After retraining the pruned model, the preserved filters reveal key insights into the feature extraction process. Fig.~\ref{fig:sinc_filters} illustrates the trained Sinc filters for sensors numbered 5 to 8. These filters reflect the optimized cluster sizes determined during pruning, resulting in a varying number of filters per layer.
    
        The top row of Fig.~\ref{fig:sinc_filters} displays the time-domain representations of the learned Sinc filters, while the bottom row shows their corresponding frequency responses. These frequency bands reveal how the pruned filters capture distinct features relevant to distinguishing healthy and patient signals.
    
        

    \subsection{Effect of filters and sensors}
    
    After training the pruned model, we focus on interpreting the feature extraction process within the SincNet layers. By studying the energy distributions of the signals passing through these layers, we aim to uncover how the model distinguishes between healthy and patient signals. This analysis emphasizes the role of the pruned Sinc filters as high-level feature extractors, effectively capturing critical frequency-based patterns for classification.
    
    We aim to calculate the signal energy from the outputs of the pruned SincNet layers, treating these as abstract processed representations of the input signals. This enables us to identify key distinctions in energy patterns between the two classes, offering insights into the discriminatory power of the filters and sensors.
    
    Initially, we seek to identify representative signals for patient and healthy classes. One intuitive approach is to compute the mean signal energy for each class. However, using the mean of all signals may incorporate noise, resulting in similar energy distributions for both classes. This similarity can obscure the distinctions necessary to identify significant sensors and critical signals that enhance classification accuracy. To address this issue, we employ clustering methods such as DBSCAN~\citep{dbscan} to identify core signals that effectively represent each class.

\begin{figure}
        \centering
        \includegraphics[width=2.75in]{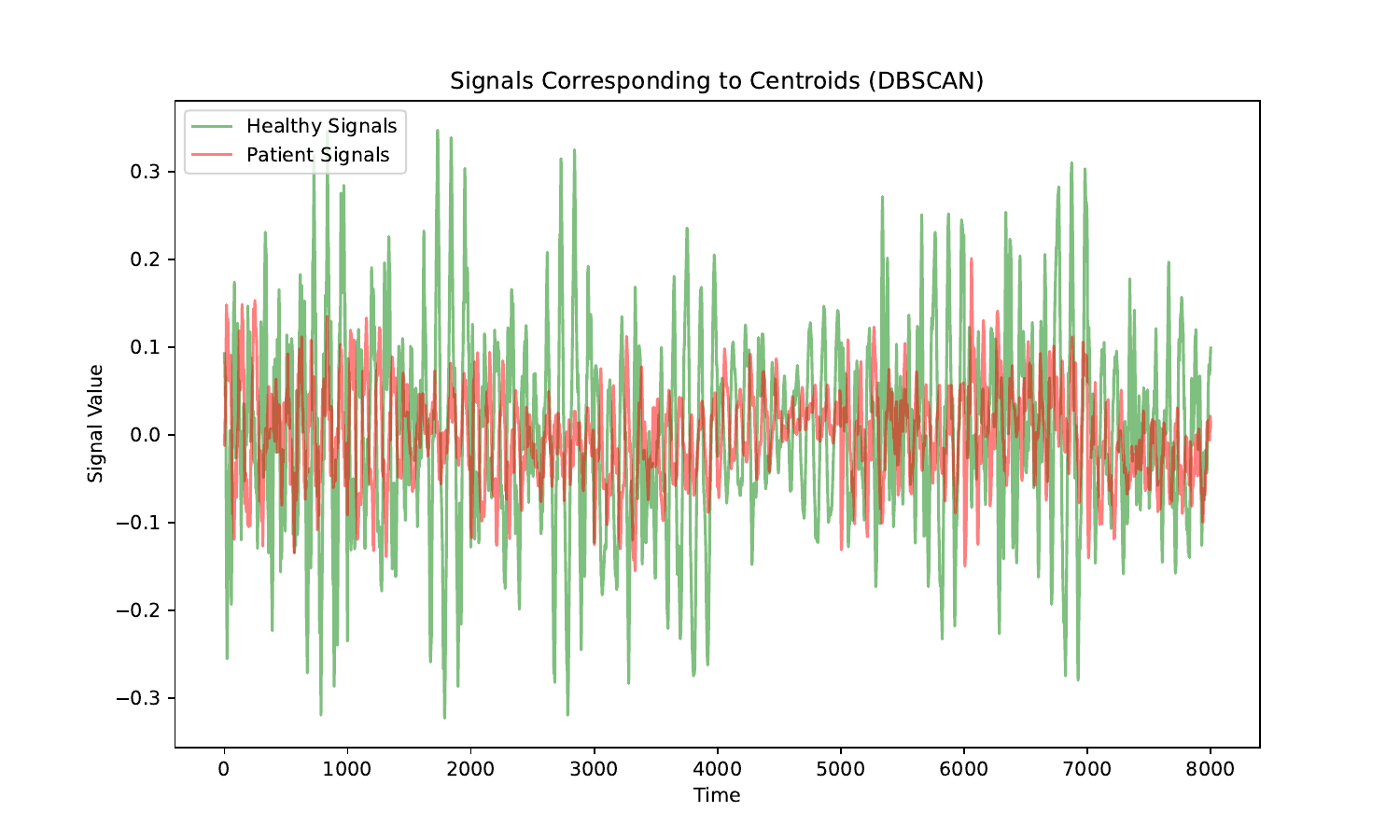}
        \caption{DBSCAN clustroids representing healthy and patient signals.}
        \label{fig:dbscan}
    \end{figure}

    By employing DBSCAN and focusing on clustroids, we identify the most representative real samples of healthy and patient signals. These clustroids serve as class representatives, enabling accurate evaluation of energy distributions and highlighting key differences between PD and healthy signals. Fig.~\ref{fig:dbscan} depicts the clustroids obtained using DBSCAN clustering.
    
\begin{figure}
        \centering
        \includegraphics[width=2.5in]{ 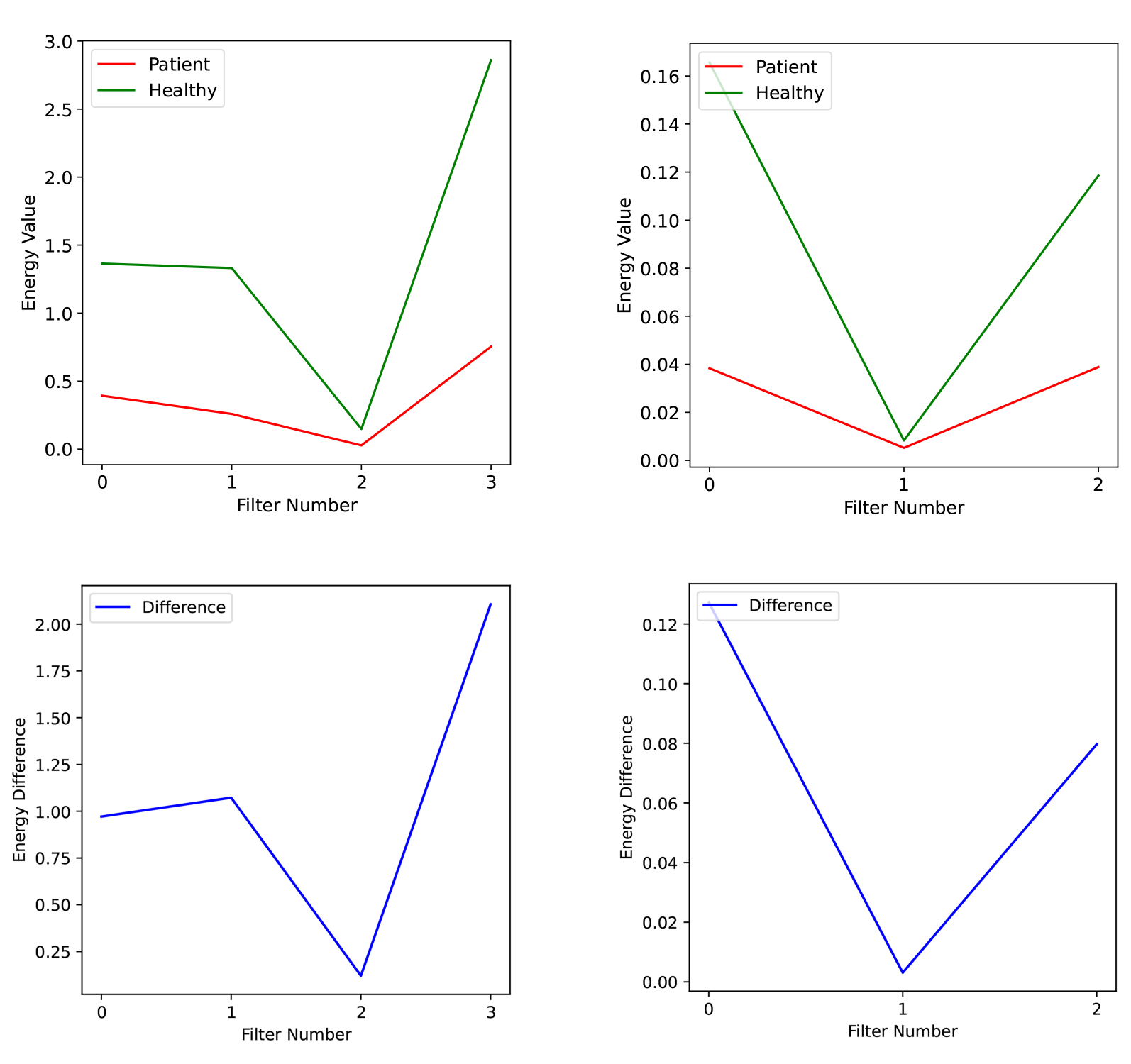}
        \caption{Energy distribution for patient and healthy signals using centroids for two sample sensors (Sensor 1 and Sensor 4, first row) and the corresponding differences in energy values across filters (second row).}
        \label{fig:energy_filters_centroids}
    \end{figure}

   Once the representative signals for patient and healthy classes are identified, they are passed through the pruned SincNet layers to analyze the output energy distributions. This analysis helps determine which specific filters within each sensor are most significant for distinguishing between the two classes. For example, in Sensor 1 (top-left corner of Fig.~\ref{fig:energy_filters_centroids}), the output energies of the first, second, and fourth filters exhibit noticeable differences between healthy and patient signals. These differences suggest that the corresponding Sinc filters are targeting specific frequency ranges crucial for classification. To systematically identify such impactful filters, we calculate the difference in output energy between healthy and patient signals across all filters of the eight sensors.
    
    As illustrated in Fig.~\ref{fig:energy_filters_centroids} (second row), the corresponding differences in output energy values for two sample sensors (Sensor 1 and Sensor 4) are shown, highlighting some filters with notable variations between the two classes.

\begin{figure}
        \centering
        \includegraphics[width=2.25in]{ 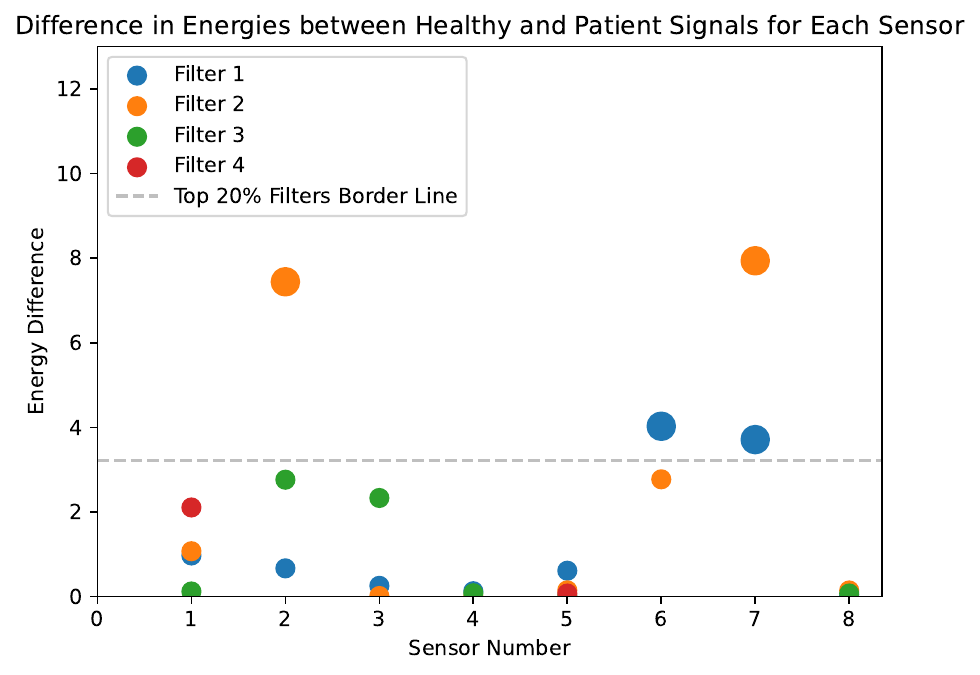}
        \caption{Distribution of Energy Differences Across Sensors and Filters}
        \label{fig:energy_difference_distribution}
    \end{figure}

    We then analyze the distribution of these energy differences across all sensors and their pruned filters to identify the discriminatory power of individual filters and sensors. Fig.~\ref{fig:energy_difference_distribution} illustrates the distribution of energy differences for all sensors and filters. From this analysis, we observe that certain sensors and filters exhibit larger energy differences, making them critical for further investigation. To better understand the characteristics of these impactful filters, we focus on analyzing the top 20\% based on their energy difference metric.
    These filters could provide valuable information for understanding the mechanisms of the model's decision-making process. likely capturing crucial features for distinguishing between healthy and patient signals. 
    
    \begin{figure}
        \centering
        \includegraphics[width=3in]{ 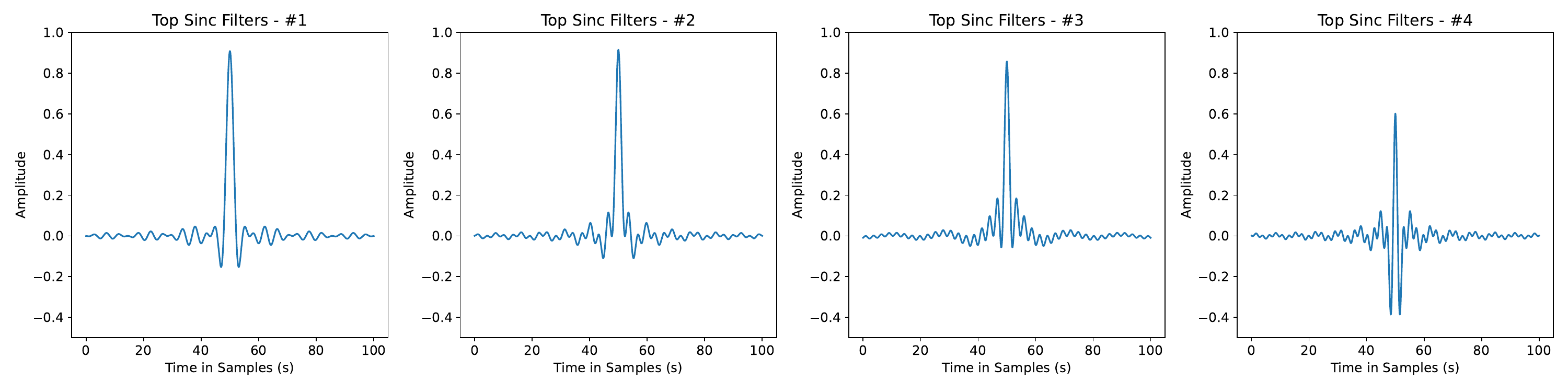}
        \caption{Top Filters Based on Energy Differences (Time Domain)}
        \label{fig:top_filters_times}
    \end{figure}
    
    \begin{figure}
        \centering
        \includegraphics[width=3in]{ 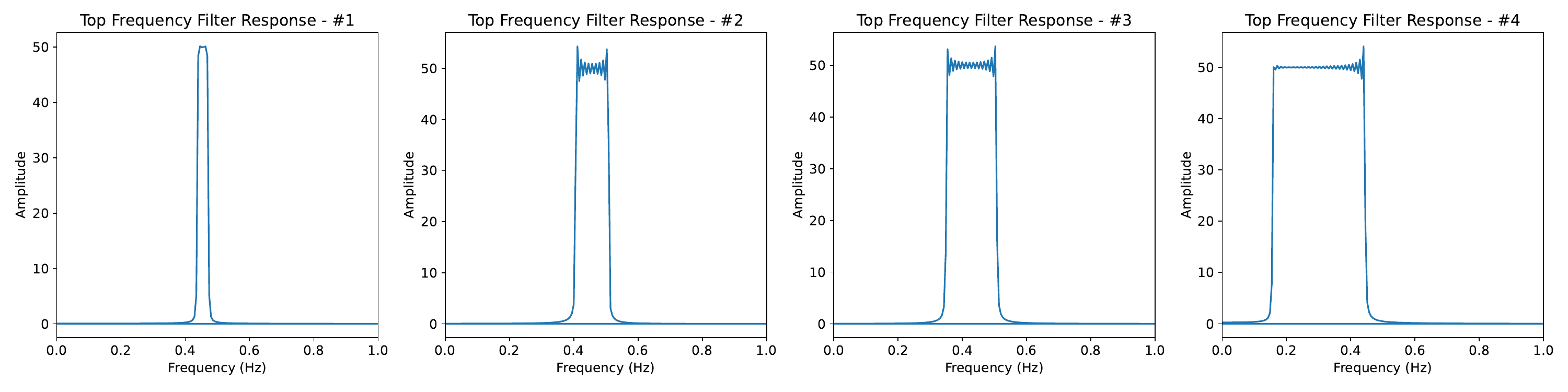}
        \caption{Top Filters Based on Energy Differences (Frequency Domain)}
        \label{fig:top_filters_freq}
    \end{figure}
    
     Fig.~\ref{fig:top_filters_times} and Fig. \ref{fig:top_filters_freq} display the top filters identified in the time and frequency domains, respectively. From Fig.~\ref{fig:top_filters_freq}, the first two top filters, primarily associated with Sensor 7 (ball of the foot) and Sensor 2 (heel), as indicated in Fig.\ref{fig:pre} for sensor locations, focus on bandpassing frequencies in the range of approximately 0.4 to 0.6 Hz while attenuating other frequencies. Similarly, the fourth filter, derived from Sensor 7, targets a narrower band of 0.2 to 0.5 Hz. These frequency ranges highlight filter selectivity in targeting class-specific features, underlining their role in distinguishing between healthy and patient gait signals.

    Furthermore, based on Fig.~\ref{fig:energy_difference_distribution}, sensors positioned at the front (e.g., Sensors 6 and 7) and back (e.g., Sensors 1 to 3) of the foot demonstrate higher energy differences. This suggests that these regions contribute more prominently to the model's feature extraction process, potentially due to their biomechanical significance in gait patterns.
    

\subsection{Severity Model Integration}

We integrate the severity model for Parkinson’s disease into our framework using transfer learning. A pretrained cluster model, obtained via the pruning method described in Section~3.5, provides the initial weights for the severity model. To preserve learned representations, we freeze the clustered model’s layers and align the severity model’s layers to these pre-trained weights. This setup leverages the feature extraction capabilities already established by the clustered model.

\begin{figure}
        \centering
        \includegraphics[width=3in]{ 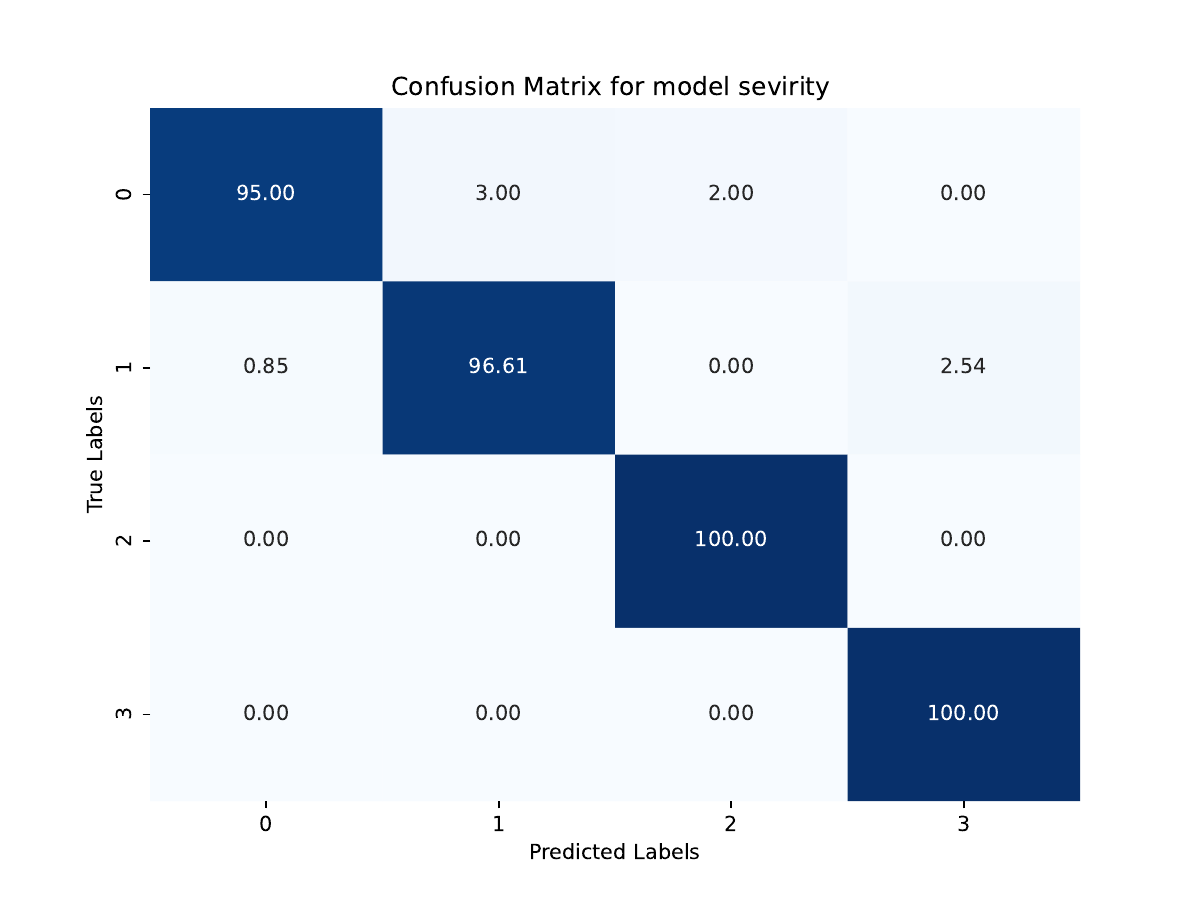}
        \caption{ Confusion matrix for severity model}
        \label{fig:confusion_matrix_Severity}
\end{figure}

After initialization, the severity model undergoes further training or evaluation to classify PD severity levels. It achieves a 97.22\% accuracy in multi-class classification. The confusion matrix for these severity predictions is shown in Fig.~\ref{fig:confusion_matrix_Severity}.

\subsection{Performance Comparison}

Our proposed model demonstrates superior performance in both binary and multi-class classification tasks compared to state-of-the-art methods, as summarized in Tables~\ref{tab:performance_comparison_binary} and~\ref{tab:performance_comparison_severity}.

\begin{table}[ht]
\scriptsize
\centering
\captionsetup{font={scriptsize,rm}}
\caption{Model Performance Comparison Before and After Pruning State of Art Methods for Binary Classification}
\resizebox{\columnwidth}{!}{
  \begin{tabular}{|c|c|c|c|c|c|}
  \hline
  \textbf{Method} & \textbf{Acc (\%)} & \textbf{Prec (\%)} & \textbf{Recall (\%)} & \textbf{F1 (\%)} & \textbf{Params} \\
  \hline
  LSTM\textsuperscript{1} & \underline{98.60} & 98.23 & 96.6 & 98.95 & \underline{1.12M} \\ \hline
  Multi LSTM\textsuperscript{2} & 91.95 & 88.75 & 96.66 & 92.53 & 33.86K \\ \hline
  CNN+LSTM\textsuperscript{3} & 98.09 & \textbf{99.22} & \textbf{100} & 98.04 & 89.2M \\ \hline
  ResNet-101\textsuperscript{4} & 97.56 & - & 97.73 & - & 44.5M \\ \hline
  \textbf{SincPD} & \textbf{98.77} & \underline{98.67} & \underline{99.55} & \textbf{99.11} & 1.17M \\
  \textit{Before Pruning} & & & & & \\ \hline
  \textbf{SincPD} & 98.15 & 98.66 & 98.66 & 98.66 & \textbf{872K} \\
  \textit{After Pruning} & & & & & \\ 
  \hline
  \end{tabular}
}
\label{tab:performance_comparison_binary}
\end{table}


\begin{table}[ht]
\scriptsize
\centering
\captionsetup{font={scriptsize,rm}}
\caption{Comparison of Model Performance with State of Art Deep Learning Methods for Multi-Classification}
\resizebox{\columnwidth}{!}{
  \begin{tabular}{|c|c|c|c|c|}
  \hline
  \textbf{Method} & \textbf{Acc (\%)} & \textbf{Prec (\%)} & \textbf{Recall (\%)} & \textbf{F1 (\%)} \\
  \hline
  1D CNN\textsuperscript{5} & 85.23 & 87.30 & 85.3 & 85.3  \\ \hline
  LSTM\textsuperscript{2} & 96.60 & \textbf{98.70} & \underline{96.20} & \textbf{97.43}  \\ \hline

  ANN-FFT\textsuperscript{6} & \underline{97.00} & \underline{98.00} & 94.00 & 96.00 \\ \hline
  \textbf{SincPD} & \textbf{97.22} & 97.30 & \textbf{97.22} & \underline{97.24}  \\ \hline
  \end{tabular}
}
\label{tab:performance_comparison_severity}
\end{table}
\footnotetext{* \textsuperscript{1} \citep{LSTM}, \textsuperscript{2} \citep{SalimiBadr-MultiLSTM}, \textsuperscript{3} \citep{Liu2021CNN_and_LSTM}, \textsuperscript{4} \citep{resnet-101}, 
\textsuperscript{5} \citep{deep1d}, 
\textsuperscript{6} \citep{ANN-FFT}.}

\noindent Tables~\ref{tab:performance_comparison_binary} and~\ref{tab:performance_comparison_severity} illustrate that the proposed model outperforms existing state-of-the-art methods in both binary and multi-class classification tasks. For binary classification, it achieves an accuracy of 98.77\% before pruning and maintains 98.15\% post-pruning with only 872K parameters, surpassing model like LSTM  in both performance and efficiency. In multi-class classification, the model attains a 97.22\% accuracy, which is competitive with top methods such as LSTM and LSTM-CNN combinations. Additionally, the proposed model offers enhanced explainability compared to existing approaches. These results demonstrate the proposed model’s superior accuracy, precision, recall, and F1 scores while significantly reducing model complexity, underscoring its effectiveness, scalability, and interpretability in diverse classification scenarios.

\section{Conclusions}
\label{sec6}
In this paper an interpretable deep structure is proposed to classify patients with Parkinson's disease and healthy subjects according to their gait cycle pattern. The proposed method can also determine the severity of the disease. 

The proposed method is a deep structure that takes the low-level raw vertical Ground Reaction Force (vGRF) signal recorded by 16 sensors put in the subjects' shoes as the input and extract higher level features based on applying various filters. Our proposed method applies bandpass filters with sinc-shape impulse responses in its first layers. Consequently, model is encouraged to learn the main frequencies of each sensor which is various between patients and healthy movements. This extraction leads to extract more meaningful features. These filters have lower number of parameters that make the method a light-weight deep model. Moreover, based on analyzing the active filters during the inference process, the important filters and sensors are determined. This information can be utilized to explain the network's output.

To train the proposed method, first a large model with a lot of filters is trained. Next, the extracted bandpass filters are studied and clustered based on their cut-off frequencies. Based on this clustering, the extra filters are pruned by replacing them with the medoids of the extracted clusters. 

The model achieved an accuracy of 98.77\% and 98.15\%  before and after applying the pruning process in PD diagnosis, and an accuracy of 97.22\% in severity detection problem. The proposed method outperforms previous models with a more parsimonious structure while providing more explanations on the reasons behind its decisions.
\newpage
\section*
\noindent\textbf{Authors' Contributions}
\textbf{Armin Salimi-Badr}: Conceptualization, Methodology, Formal analysis, Theoretical analysis, Validation, Investigation, Supervision, Writing - original draft, Writing – review \& editing, Project administration.
\textbf{Mahan Veisi} and \textbf{Sadra Berangi}: Methodology, Software, Validation, Data curation, Visualization, Writing - original draft. 

\noindent\textbf{Funding}
The is not any funding .

\noindent\textbf{Data Availability}  The dataset used during the current study is an open access dataset from free public database, available in PhysioNet\footnote{https:$\slash$$\slash$physionet.org$\slash$content$\slash$gaitpdb$\slash$1.0.0$\slash$}. 

\section*{Declarations} 
\textbf{Conflict of Interest}  None.

\noindent \textbf{Ethical Approval} Not applicable.

\noindent\textbf{Consent to Participate} Not applicable.

\noindent\textbf{Consent for Publication} Not applicable.

\noindent\textbf{Clinical Trial Number} Not applicable.

\setlength{\bibsep}{0pt} 
\bibliography{myref.bib}

\end{document}